\documentclass[12pt,preprint]{aastex}
\def\edth{\;\raise1.0pt\hbox{$'$}\hskip-6pt\partial\;}
\def\baredth{\;\overline{\raise1.0pt\hbox{$'$}\hskip-6pt
\partial}\;}
\def\gsim{~\rlap{$>$}{\lower 1.0ex\hbox{$\sim$}}}
\def\lsim{~\rlap{$<$}{\lower 1.0ex\hbox{$\sim$}}}

\def\d{{\rm d}}

\def\PRD{Phys. Rev. D}

\def\be{\begin{equation}}
\def\ee{\end{equation}}
\def\bea{\begin{eqnarray}}
\def\eea{\end{eqnarray}}

\slugcomment{submitted to ApJ}
\begin{document}

\title{Primordial quadrupole-induced polarisation from filamentary
  structures and galaxy clusters} \author{Guo-Chin
  Liu\altaffilmark{1,2}, Antonio da Silva\altaffilmark{1} \& Nabila
  Aghanim\altaffilmark{1} }
  \altaffiltext{1}{IAS-CNRS, B\^atiment 121, Universit\'e Paris Sud,
  91405, Orsay} \altaffiltext{2}{Institute of Physics, Academia
  Sinica, Taipei, Taiwan}
\begin{abstract}
We present the first computation of the Cosmic Microwave Background
(CMB) polarisation power spectrum from galaxy clusters and filaments
using hydrodynamical simulations of large scale structure.  We give
the $E$ and $B$ mode power spectra of the CMB quadrupole induced
polarisation between $\ell \sim 560$ and $20000$. We find that the
contribution from warm ionised gas in filamentary structures dominates
the polarised signal from galaxy clusters by more than one order of
magnitude on large scales (below $\ell \sim 1000$) and by a factor of
about two on small scales ($\ell \gsim 10000$). We study the
dependence of the power spectra with $\sigma_8$. Assuming the power
spectra vary like $\sigma_8^n$ we find $n=3.2-4.0$ for filaments and
$n=3.5-4.6$ for clusters.

\end{abstract}
\section{Introduction}

After the success of the extraction of cosmological
information from the Cosmic Microwave Background (CMB) temperature
anisotropies, much effort is now put towards building experiments for
measuring the CMB polarised signal, which is about one order of
magnitude smaller.
There are many scientific goals of the CMB polarisation studies. These
include determining the reionisation epoch from the large scale
polarisation power spectrum, testing inflationary models by searching
specific patterns due to the existence of the primordial gravitational
waves, and improving the precision of cosmological parameters derived
by CMB temperature anisotropies alone.  Aiming at these objectives,
the satellites WMAP~\citep{wmap} and Planck~\citep{planck},
Balloon-borne experiments like ARCHEOPS~(http://www.archeops.org/),
Boomerang~\citep{boom} and MAXIPOL~\citep{max} and Ground-based
instruments like AMiBA~\citep{AMIBA} and DASI~\citep{dasi} have observed
or will observe the polarised sky.

An important issue concerning the measurement of the primordial
polarisation power spectrum is that it is not the only polarised
source in the sky. In addition to the contamination from the so-called
foregrounds such as galactic dust~\citep{benoit}, free-free and
synchrotron emissions~\citep{Oliveira}, there are other sources of
polarisation due to large scale structure (LSS)
(reionisation~\citep{ng}, weak lensing~\citep{ZS2}, ...).  In
particular, when the CMB photons pass through the hot ionised gas
trapped in the potential wells of galaxy clusters a polarised signal
is induced.  The presence of the CMB temperature quadrupole induces a
linear polarisation in the scattered radiation.  Sunyaev and
Zel'dovich (1980) were the first to estimate the level of such a
polarisation induced by galaxy clusters. These authors pointed out
that in addition to the primary CMB quadrupole there are two other
sources of temperature quadrupole seen by a cluster: A quadrupole due
to the transverse peculiar velocity of the cluster, and double
scattering. Studying the induced CMB polarisation due to clusters can
open a whole new window for cosmology. Sunyaev and Zel'dovich proposed
to use the polarisation to estimate cluster's transverse
velocity. Audit \& Simmons (1999) gave a description of the effect due
to the kinetic quadrupole, including the frequency
dependence. Moreover, measuring the polarisation towards distant
clusters should in principle permit to provide to observe the
evolution of the CMB quadrupole. The CMB quadrupole seen by clusters
contains statistical information on the last scattering surface at the
cluster position.  Therefore, measuring the cluster polarisation
should help us to beat the cosmic variance~\citep{KL,JP}.  Sazonov \&
Sunyaev (1999) revisited the detailed polarisation signal due to the
three sources of quadrupole given the constraint on the CMB quadrupole
by COBE~\citep{bennett}, the electron density profiles of clusters and
their transverse peculiar velocities. For the COBE data, their
sky-average polarisation induced by the primordial quadrupole was
found to be $3.1\tau\mu{\rm K}$, with $\tau$ the cluster optical
depth. Additionally, if clusters possess an intra-cluster magnetic
field, Faraday rotation will affect the linearly polarised CMB. This
effect was investigated by Takada et al. (2001) who found an amplitude
of the order of the microgauss at low frequencies. Inversely, the
effect of Faraday rotation on the CMB polarisation was proposed to
extract the information of the cluster magnetic field when the cluster
electron density is obtained from SZ and X-ray
observations~\citep{Ohno}.

In this paper we concentrate on the case where no magnetic field is
present in the intra-cluster gas. We study the polarised signal due to
LSS by calculating the polarised angular power spectrum.  Several
authors have done some research in the polarisation due to ionised gas
in clusters or at the reionisation. Hu (2000) investigated the
secondary polarisation induced by several physical processes using an
analytic method. For the polarisation in galaxy clusters, the author
considered a density modulation of the kinetic and primordial
polarisation sources similar to the Vishniac
effect~\citep{Vishniac}. Liu et al. (2001) computed the polarised
signal at reionisation using N-body simulations in combination with an
analytic description to model the gas distribution.  More recently,
Santos et al. (2003) investigated analytically the contribution from
ionised patches at reionisation. They assumed the ionising sources
reside in the dark matter halos and used the dark matter correlations
in linear approximation. Both latter papers deal with the a
polarisation induced by the primordial CMB quadrupole since it
typically dominate over the kinetic and double scattering contribution
at scales of interest.  Concerning galaxy clusters, Baumann et
al. (2003) computed the polarisation induced by primordial and kinetic
quadrupole using a halo-clustering approach to describe fluctuations
in the electron density. Lavaux et al. (2004) focus on the polarised
signal due to the kinetic quadrupole and the double scattering by the
use of hydrodynamical simulations.

In the present work, we revisit the polarisation signal induced by
large scale structure. We use the analytical method developed in Liu
et al. (2001) combined for the first time with state-of-the-art
hydrodynamical simulation techniques.  We concentrate on the the
primordial CMB quadrupole induced polarisation as it is the dominant
source. Hydrodynamical simulations are the most powerful tool to
describe the physics of the gas. Taking advantage of this, we
investigate the contribution to the polarised signal of different gas
phases, namely the hot gas in galaxy clusters and the warm gas in
filaments. In Section 2 and 3 we describe the method and
the hydrodynamical simulations we used for our estimate. In Section 4,
we present the results of our work and give a discussion and
conclusion in Section 5 and 6 respectively.

Throughout this paper we will assume a flat $\Lambda$CDM cosmology with
cosmological parameters matching present observations:
matter density, $\Omega_{{\rm m}}=0.3$, cosmological constant density,
$\Omega_{\Lambda}=0.7$, baryon density, $\Omega_{\rm b}=0.044$, and Hubble
parameter, $h=0.71$. 

\section{Background}

\subsection{Analytic formulae}

The polarised CMB signal is usually described by two of the Stokes parameters
${\cal Q}$ and ${\cal U}$. 
If we consider a wave traveling in the $\hat{z}$ direction, ${\cal Q}$
is the difference in intensity in the $\hat{y}$ and $\hat{x}$
directions, while ${\cal U}$ is the difference in the
$(\hat{x}+\hat{y})/\sqrt{2}$ and $(\hat{x}-\hat{y})/\sqrt{2}$
directions.  The circular polarisation parameter, ${\cal V}$, cannot
be produced by scattering, so we will not mention it further.
Polarisation calculations are more complicated than temperature
calculations because the values for ${\cal Q}$ and ${\cal U}$ depend
on the choice of coordinate system.  Therefore, it is more convenient
to decompose the polarisation on the sky into a divergence free
component, the so-called $E$ mode, and a curl component, the so-called
$B$ mode, which are coordinate independent.

We are interested in computing the $E$ and $B$ mode polarisation power
spectra due to galaxy clusters and filaments. For this, we use the
analytic formulae detailed in Liu et al. (2001) to which we refer the
reader. In this work, the electron density fluctuation modulated
quadrupole at any given scale ${\bf k}$ and conformal time $\eta$ is
\begin{eqnarray}
S^{(m)}({\bf k},\eta)&\equiv& \int \d^3 {\bf p}\delta_{\rm e}
({\bf k-p},\eta)\Delta_{T2}^{(m)}({\bf p},\eta)\nonumber \\
&\simeq&
\delta_{\rm e}({\bf k},\eta)\int \d^3 {\bf p}\Delta_{T2}^{(m)}
({\bf p},\eta)\nonumber \\
&\equiv& \delta_{\rm e}({\bf k},\eta)Q^{(m)}_2(\eta),
\label{defS}
\end{eqnarray}
where $\delta_e$ is the electron density fluctuation,
$\Delta_{T2}^{(m)}$ is the primordial CMB quadrupole with angular
momentum $m= 0, \pm 1, \pm 2$ and $Q^{(m)}_2(\eta)\equiv \int \d^3
{\bf p}\Delta^{(m)}_{T2}({\bf p},\eta)$. The approximation in the
second line assumes that the first order temperature quadrupole
$\Delta^{(m)}_{T2}$ is uncorrelated with $\delta_{\rm e}$. In
other words, the dominant contributions to the modulated quadrupole
come mainly from the CMB quadrupole at large scales and from
the electron density fluctuations at small scales, see Hu (2000).  The
temperature anisotropy field produced by scalar mode perturbations
has axial symmetry. Therefore, the quadrupole field is decomposed into
the $m=0$ component in the ${\bf p}$-basis parallel to the axis of
symmetry, i.e. the temperature anisotropy quadrupole is
$\Delta^{(0)}_{T2}Y_2^0(\beta,\alpha)$, where $\beta$ and $\alpha$ are
the polar and azimuthal angles defining ${\bf \hat n}$ in this
basis. Using the addition theorem, we can project the component in the
${\bf p}$-basis onto the ${\bf k}$- basis (see Ng \& Liu 1999),
\begin{equation}
\sum_{m}Y^{m*}_\ell({\bf \hat{n}})Y_\ell^m({\bf
\hat{p}})=\sqrt{\frac{2\ell+1}{4\pi}}Y_\ell^0 
(\beta,\alpha).
\end{equation} 
It then follows that
\begin{equation}
Q^{(m)}(\eta)=\sqrt{\frac{4\pi}{5}}\int\d^3 {\bf p} \Delta^{(0)}_{T2}
({\bf p},\eta)Y_2^{m*}(\hat{\bf p}).
\label{Qm}
\end{equation}

Provided the modulated quadrupole $S$ Eq.~(\ref{defS}), the polarised
power spectrum can be obtained by integrating the Boltzmann equation
along the photon past light cone. Here we skip the detailed
derivation given in Liu et al., the power spectra for the $E$ and
$B$ modes are:
\begin{equation}
C_{(E,B)\ell}=(4\pi)^2\frac{9}{16}\frac{(\ell+2)!}{(\ell-2)!} \sum_m\int
k^2 \d k \left\langle \left |\int \d\eta g(\eta) S^{(m)}({\bf k},\eta)
T_{(E,B)\ell}^{(m)}(kr)\right |^2 
\right\rangle.
\label{DEL2}
\end{equation}
In this expression the first term is the visibility function
$g(\eta)$, which is the probability that a photon had its last scattering at
epoch $\eta$ and reaches the observer at the present time,
$\eta_0$. It is defined as
\begin{equation}
g(\eta)\equiv
-\frac{\d\tau}{\d\eta}{\rm e}^{\tau(\eta_0)-\tau(\eta)},
\label{eqvis}
\end{equation}
where $\tau(\eta) \equiv \int_\eta^{\eta_0}d\eta' a n_{\rm e}
\sigma_{\rm T}$ is the Thomson electron-scattering optical depth at
time $\eta$.  $S^{(m)}$ represents the source term of polarisation in
Eq.~(\ref{DEL2}).  The last quantity is a geometrical factor
$T_{(E,B)\ell}^{(m)}(kr)$ with $r=c(\eta_0 -\eta)$ and $c$ the speed
of light. It is given by the combination of the spherical Bessel
functions. We list the results of $T_{(E,B)\ell}^{(m)}(kr)$ in
Table~1.  Note that the expression $T_{(E,B)\ell}^{(m)}(kr)$ in the
Table are valid only in flat universes. The hyperspherical Bessel
functions should be used in curved universe~\citep{Hu2}.

\begin{table}[t]
\begin{tabular}{|l|c|c|} \hline
$m$ & $T_{E\ell}^{(m)}$ & $T_{B\ell}^{(m)}$ \\ \hline $0$ &
$(-i)^\ell\frac{j_\ell(kr)}{(kr)^2}$ & 0 \\ \hline $\pm 1$ & $\mp
(-i)^\ell\frac{1}{(2\ell+1)kr}\sqrt{\frac{1}{6\ell(\ell+1)}}
[\ell j_\ell(kr)-(\ell+1)j_{\ell-1}(kr)]$ & $\pm \sqrt{\frac{3(-i)^\ell}{2\ell(\ell+1)}}
\frac{j_\ell(kr)}{(kr)^2}$ \\ \hline $\pm 2$ & $\pm
\frac{(-i)^\ell}{2\ell+1}\sqrt{\frac{(\ell-2)!}{6(\ell+2)!}}
\left(\left[\frac{(\ell+2)(\ell+1)}{2\ell-1}+\frac{\ell(\ell+1)}{2\ell+3}
+6\frac{(2\ell+1)(\ell-1) (\ell+2)}{(2\ell-1)(2\ell+3)}\right] \right.
 $ & $\mp \frac{(-i)^\ell}{2\ell+1}
\sqrt{\frac{(\ell-2)!}{6(\ell+2)!}}\left[(\ell+2)j_{\ell-1}(kr) \right.$ \\
 &$\left.\times j_\ell(kr) -(\ell+2)(\ell+1)\frac{j_{\ell-1}
(kr)}{kr}-\ell(\ell-1)\frac{j_{\ell+1}(kr)}{kr}\right )$ & $ \left .
-(\ell-1)j_{\ell+1}(kr) \right]$ \\ \hline
\end{tabular}
\caption{Contributions to the geometrical factor $T_{(E,B)\ell}^{(m)}$
for different $m$'s used in equation ({\ref{DEL2}}).}
\end{table}

\subsection{Simulation details}

From Eq.~(\ref{DEL2}) we easily see that the polarisation amplitude
depends on the density of the electrons which scatter the CMB photons
through $g(\eta )$ and $S^{(m)}$. We therefore expect that 
different gas phases contribute differently to the overall signal.
One important step of our study is thus to quantify these
contributions, using  hydrodynamical N--body simulations to model the
gas dynamics.

The simulations were generated with the public version of the {\tt
Hydra} code~(Couchnman, Thomas, \& Pearce 1995;Pearce \& Couchman
1997), which implements an adaptive particle-particle/particle-mesh
(AP$^3$M) algorithm \citep{couchman:1991} to calculate gravitational
forces and smoothed particle hydrodynamics (SPH) \citep{monaghan:1992}
to estimate hydrodynamical quantities.
In the present study, we consider a {\it non-radiative} 
model, i.e. the gas component evolves under the action of gravity,
viscous forces and adiabatic expansion. The effects of
non-gravitational heating and radiative dissipation of energy, will be
addressed in a subsequent study.

We present results from three simulation runs of the $\Lambda$CDM
cosmology adopted in this paper (see Section 1).  The initial density
and velocity fields were constructed with $160^3$ particles of both
baryonic and dark matter, perturbed from a regular grid of fixed
comoving size $L=100 h^{-1} {\rm Mpc}$, using the Zel'dovich
approximation. With this choice of parameters the mass of each
baryonic and dark matter particles were $m_{\rm gas}=2.6 \times 10^{9}
\, h^{-1} M_{\odot}$ and $m_{{\rm dark}} = 2.1 \times 10^{10} \,
h^{-1} M_{\odot}$, respectively.  We assumed a matter power spectrum
described by the CDM transfer function in Bardeen et al. (1986) (known
as the BBKS transfer function for CDM models) with shape parameter,
$\Gamma $, given by the formula in Sugiyama (1995). The
normalisation of the matter power spectrum in each simulation run was
set such that the present day r.m.s matter fluctuations in spheres of
$8h^{-1}$ Mpc radius, $\sigma_8$, was equal to 0.8, 0.9 and 1
respectively.  We used the same realisation of the power spectrum in
each run to permit direct comparison between forming structures in the
three simulations.  The runs were started at redshift $z=49$. The
gravitational softening was fixed at $25\,h^{-1} {\rm kpc}$ in
physical units between redshifts zero and one, and held constant above
this redshift to $50\,h^{-1} {\rm kpc}$ in comoving coordinates.

\section{Method}

\begin{figure}
\plotone{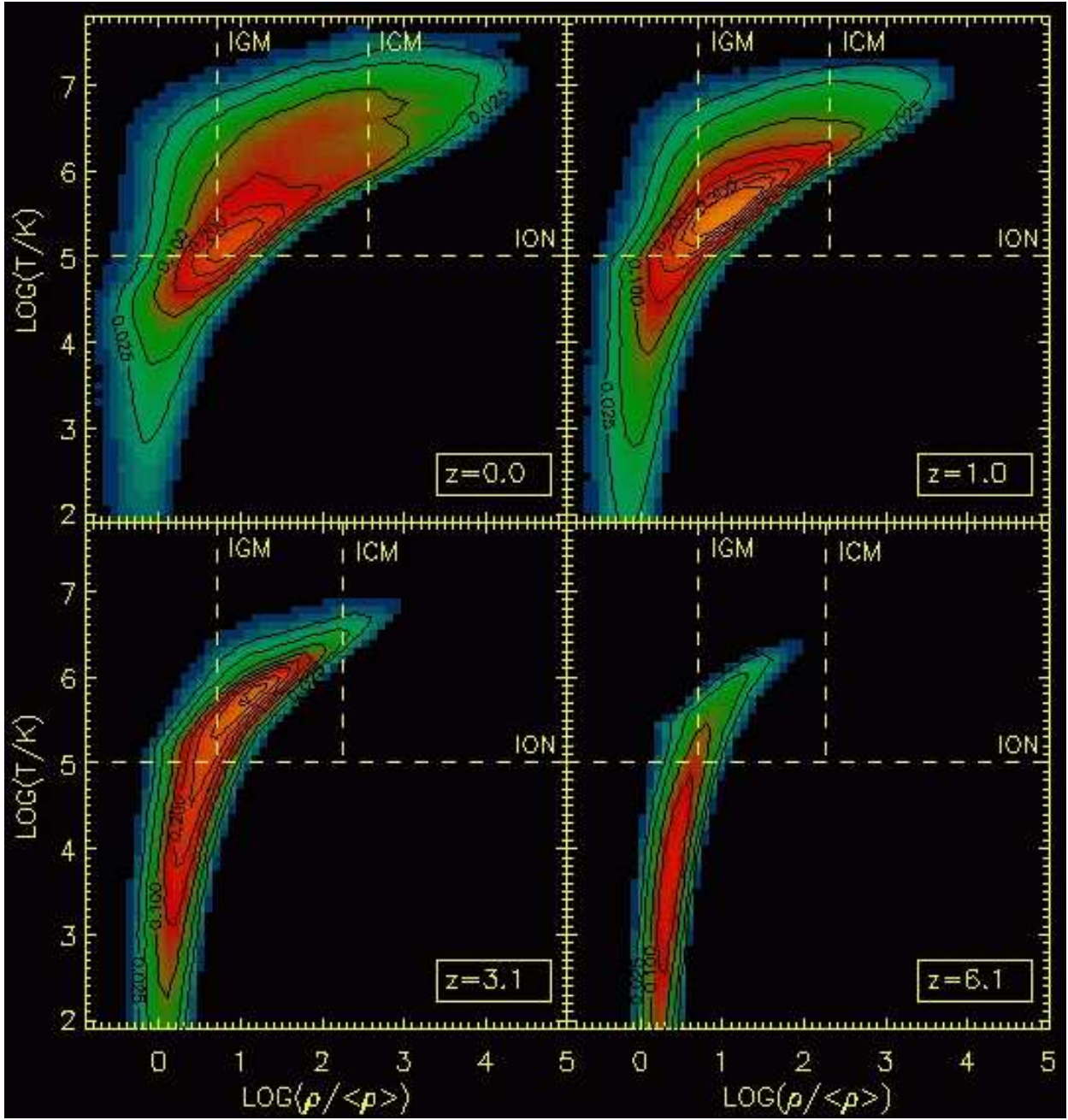}
\caption{Temperature-overdensity distribution (phase-space) of  gas
particles in the $\sigma_8=0.9$ simulation at redshifts z=0, 1, 3.1,
and 6.1 (top right, top left, bottom left and bottom right panels,
respectively). In each panel, horizontal dashed lines represent the
temperature threshold, $T_{\rm th}$, above which the gas is considered
to be ionised. The ICM and IGM phases are defined in each panel by the
vertical dashed lines.
}
\label{phase}
\end{figure}

We study the relative contribution to the polarisation power spectrum
from clusters (hot high-density gas) and filamentary structures (warm
low-density gas). We define the gas phases based on the baryon
collapsed density and temperature. We call hot ionised gas the phase
consisting of all gas particles with temperatures above the threshold
$T_{\rm th}=10^5$ K. Gas above this temperature threshold is assumed
fully ionised and is considered in the calculations as the only gas
scattering the CMB photons.  Hot ionised gas particles with
overdensities, $\delta_i$, larger than the density contrast at
collapse, $\Delta_c =178\Omega_m(z)^{-0.55}$ (Eke, Navarro, \& Frenk
1998), are turn considered to be part of the intra cluster medium
(ICM) high-density phase. Low-density ionised gas particles with
overdensities ranging in the interval $5<\delta_i<\Delta_c$ encompass
a warm intergalactic medium (IGM) phase (see e.g. Valageas,
Schaeffer, \& Silk (2002)) which we call ``filamentary structures''.
Fig.~({\ref{phase}}) shows the overdensity-temperature distribution
(phase space) of all gas particles in our $\sigma_8=0.9$ simulation
run at redshifts $z=0$ (top left panel), $z=1$ (top right panel),
$z=3.1$ (bottom left panel) and $z=6.1$ (bottom right panel). In each
case the phase space was discretised in bins of size dex=0.1 per
logarithmic decade, that gives the gas fraction density (gas fraction
per dex$^2$) in each bin. The horizontal dashed lines represent the
temperature threshold, $T_{\rm th}$, above which the gas is considered
to be ionised. The ICM and IGM phases are defined in each panel by the
vertical dashed lines.  We should note that our simulations do not
include photoinisation heating due to UV background radiation, which
explains the significant fraction of very cold low-overdensity gas.
In our present simulations, resolution together with a simplified
physical model prevent us from properly describing this gas phase.

In order to calculate the $E$ and $B$ power spectra using
Eq.~(\ref{DEL2}), we need to evaluate the electron density,
$\delta_{\rm e}({\bf k},\eta)$, and the primordial CMB quadrupole,
$\Delta_{T2}^{(m)}$ which appears in Eq~(\ref{Qm}). We modify the
publicly available code CMBFast~\citep {ZS} to output the time
evolution of the primordial CMB quadrupole at different ${\bf k}$
modes. This is then integrated in {\bf k}-space at each epoch to
obtain $Q^{(m)}_2(\eta)$ in Eq.~(\ref{defS}).  The next step is to
compute the visibility function $g(\eta)$ and the polarisation source
term Eq.~(\ref{defS}) for which the electron density fluctuations
$\delta_{\rm e}({\bf k},\eta)$, are needed.  The first quantity is
obtained directly from our simulation runs by evaluating the
ionisation fraction at different redshifts.  For the electron density
fluctuations, we start by evaluating this in real space, $\delta_{\rm
e}({\bf r},\eta)$, on a regular grid with $N_{\rm grid}^3$ cells
inside the comoving box, making use of the SPH formalism
\citep{monaghan:1992}. In this step we assume a primordial gas
composition with 24\% helium abundance.  The computation of
$\delta_{\rm e}({\bf k},\eta)$ follows by performing the Fourier
transform of $\delta_{\rm e}({\bf r},\eta)$.

\section{Power spectra of polarisation of clusters and filaments}

\begin{figure}
\plotone{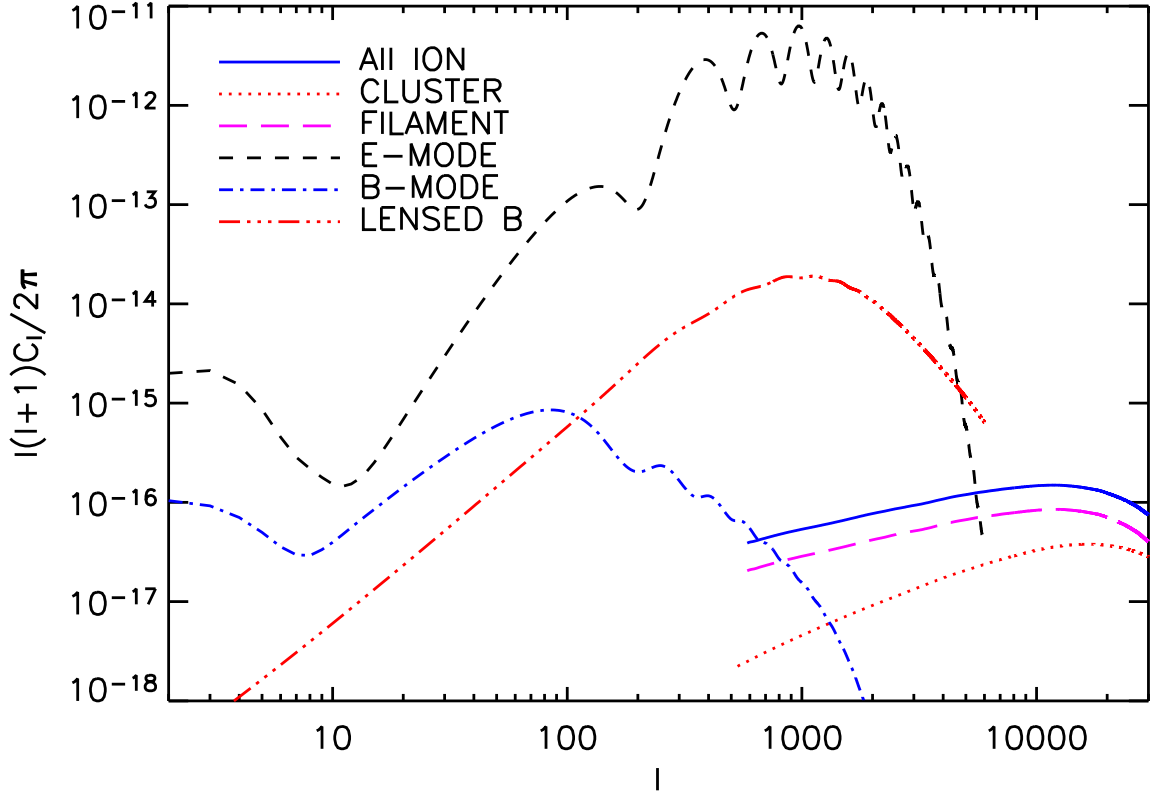}
\caption{Polarisation power spectra for $\sigma_8=0.9$. The
polarisation power spectra for all ionised particles, filaments and
clusters are shown by solid, long-dashed and doted curves,
respectively. We also plot the primary $E$ mode induced by scalar
perturbations (short-dashed line), primary $B$ mode (dot-dashed line)
from tensor perturbations and $B$ mode from lensed $E$ mode (dot-dot-dashed
line).}
\label{cl_all}
\end{figure}

We compute the $E$ and $B$ power spectra for filaments, galaxy
clusters and all ionised gas. Results obtained adopting a regular grid
with $600^3$ cells (corresponding to a fixed cell separation of
$0.17\, h^{-1} {\rm Mpc}$ in comoving coordinates) are shown in
Fig.~({\ref{cl_all}}) in the $\sigma_8=0.9$ case.  We tested the
convergence of our results by comparing $C_{(E,B)\ell}$ evaluated with
$300^3$, $450^3$ and $600^3$ grids and found no significant difference
between the last two values. This indicates we reached numerical
convergence. The domain of validity of our results is $560\lsim \ell
\lsim 20000$, outside which we reach numerical resolution limits. We
also tested for the temperature threshold. Lowering $T_{\rm th}$ to
$10^{4}$K, decreases the power spectra by no more than 20\%.

We note that we get the same power for the secondary $E$ and $B$ mode
polarisation with a maximum relative difference smaller than
$10^{-6}$. The reason for this equality is essentially that the
first-order quadrupole, whose scattering produces the polarisation, is
dominated by large scales and has thus a random orientation relative
to the small-scale electron density fluctuations. Scattering of the
quadrupole by the small-scale fluctuations therefore, on average,
excites $E$ and $B$ modes equally (e.g. Hu 2000, Liu et al. 2001).

We now compare the contributions to the polarisation signal from
clusters and filaments, the dotted and the long dashed lines,
respectively, in Fig.~({\ref{cl_all}}). For the filaments, the shape
of the power spectrum is very similar to that induced by all ionised
particles, but the amplitude is smaller by a factor of about $1.7$.
On large scales ($\ell \sim 1000$) the power spectrum of filaments
dominates that of clusters by one order of magnitude. On small scales
($\ell \sim 17000$), this difference is only a factor of two
indicating that clusters contribute mainly at large $\ell$'s, as
expected.

\begin{figure}
\plotone{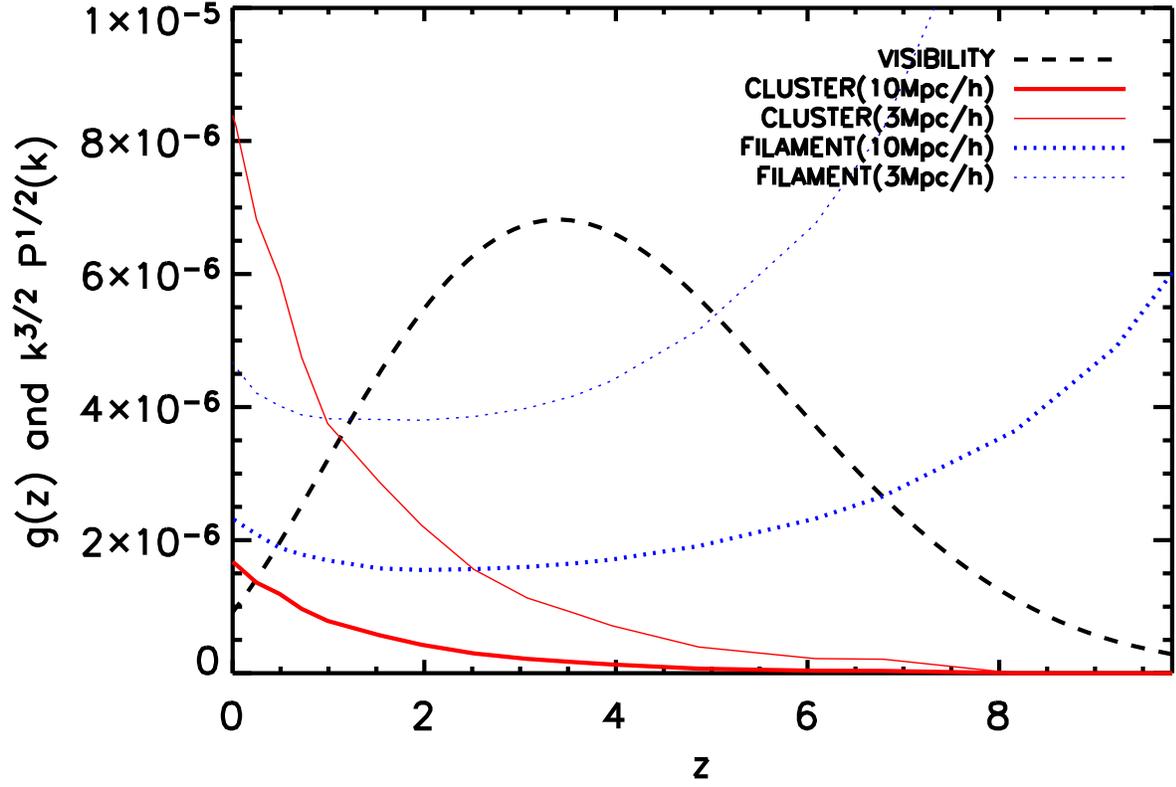} 
\caption{ Visibility function (dashed line) and square root of the
  logarithmic power for the clusters and filaments phases at two
  characteristic scales (3 and 10 $h^{-1}$Mpc). Results are displayed
  using an arbitrary normalisation factor for the power.}
\label{VIS}
\end{figure}

We illustrate the difference between cluster and filament polarised
power spectra by plotting in Fig.~(\ref{VIS}) the evolution of the
visibility function (thick dashed line) and density perturbations at
two typical scales 3 and 10 $h^{-1}{\rm Mpc}$ for clusters and
filaments, respectively.  Recall the Eq.~(\ref{DEL2}), the power
spectrum is the integral of the electron density weighted by the
visibility function. At the typical cluster scale, the power does not
exceed that of filaments, down to $z \sim 1$. At that redshift, the
visibility function from our simulation is relatively small. It peaks
at $z \gsim 3.2$ and its full width half maximum is bounded by $z \sim
1.1$ and $z \sim 6.2$. In this range of redshifts, the power from
filaments is much larger than from clusters. As a consequence, most of
the scattering producing the linear polarisation occurs for $z \gsim
1.2 $ when the relative contribution of clusters is still small.

The power spectrum of polarisation induced by total ionised particles
(solid curve in Fig~({\ref{cl_all}})), peaks at a scale of about $\ell
\sim 10000$ with an amplitude much smaller than the primary $E$ mode
polarisation.  We now compare the power spectrum of the secondary
polarisation and the power spectrum of the secondary temperature
anisotropies due to SZ effect for the all ionised phase. We base our
comparison on the results of da Silva et al.~\citep{antonio}.  They
have estimated the SZ temperature power spectrum using similar
non-radiative simulations. The amplitude of the SZ temperature power
spectrum is five orders of magnitudes larger than the polarisation
signal (see Fig.~3 in da Silva et al. (2001)). In Fig.~4 of da Silva
et al. (2001), the power spectrum of temperature anisotropies
associated with clusters, in the non-radiative case, dominates by
almost a factor 2 that of filaments whereas the power spectrum of
polarised signal is dominated by filaments. This is because the
amplitude of the SZ temperature anisotropy is an integral of the free
electron density weighted by the gas temperature. Typically,
the temperature in clusters is one hundred times higher than the
temperature in filaments. The secondary polarisation signal, although
weak, can provide complementary information to that we get from
secondary temperature fluctuations especially for the filaments.

\section{Discussion}

This is the first time we are able to investigate simultaneously the
polarisation contribution from ICM and IGM gas phases using
hydrodynamical simulations. Therefore direct comparisons with previous
studies, based either on analytic computations or on N-body
simulations, are not straightforward especially for the filamentary
structures but also for clusters. As an example, our angular power
spectrum for cluster matches the one given by Santos et
al. (2003) particularly below $\ell \sim 1000$. However
they consider, in their halo model, structures within
$100<T_{vir}<10^4$K which we do not model here (see Sec. 2.2).  Our
results from the ICM phase are easier to compare by nature with those
from Liu et al. (2001), which are based on N-body computations.  Our
polarisation power spectrum for clusters is in very good agreement
with the relevant models, namely A and B, in Fig.~4 of Liu et
al. Finally, we find that the amplitude of the power spectrum induced
by galaxy clusters, at its maximum, is consistent with the estimate of
Sazonov \& Sunyaev~ (1999) for the sky-averaged polarisation:
$3.1\tau\mu{\rm K} \sim 30\ {\rm nK}$, with a typical optical depth for
clusters $\tau=0.01$. From our hydro-dynamical simulation we find
an average value of 24 nK.

\begin{figure*}
\includegraphics[angle=0,width=0.5\linewidth]{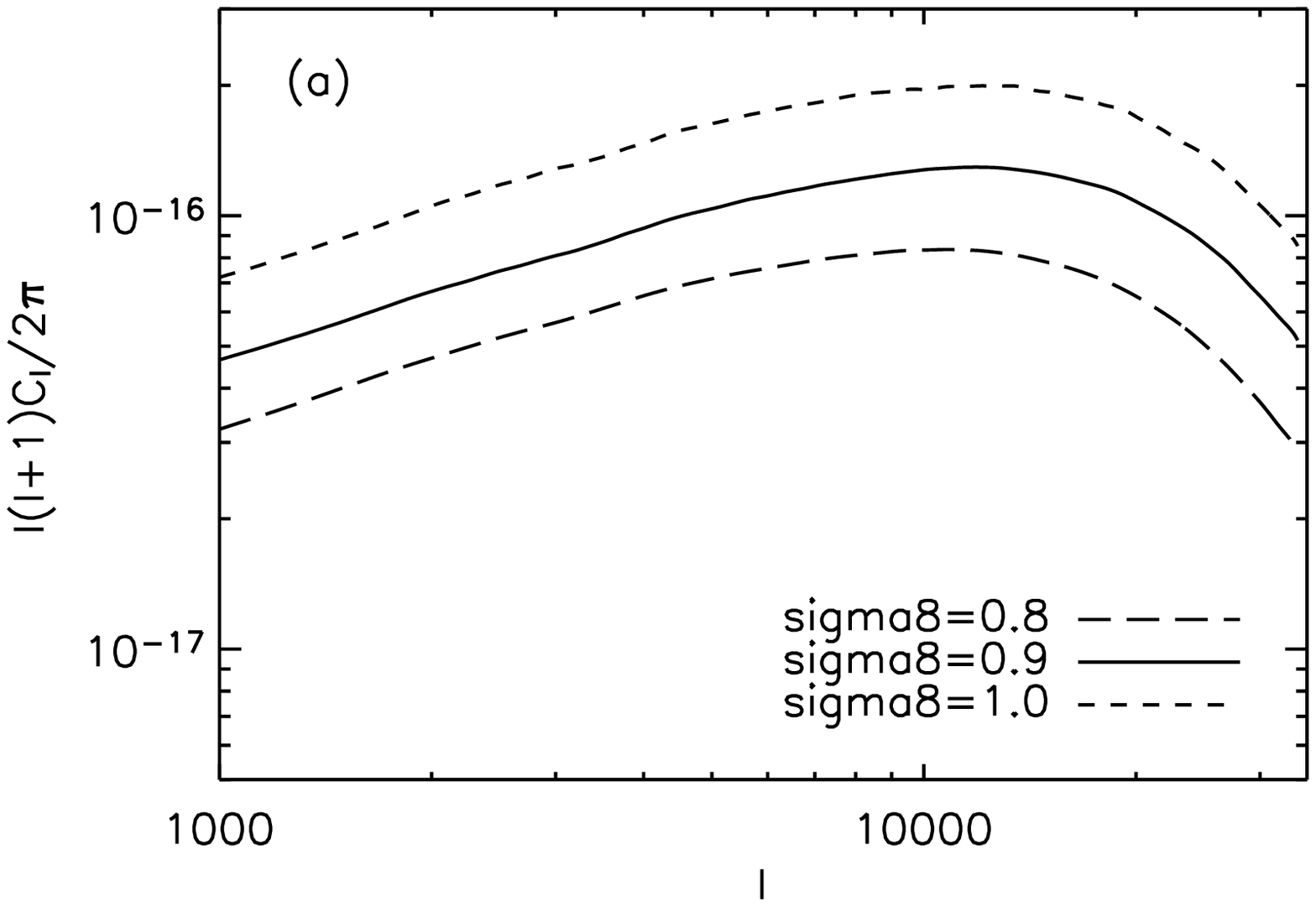}
\includegraphics[angle=0,width=0.5\linewidth]{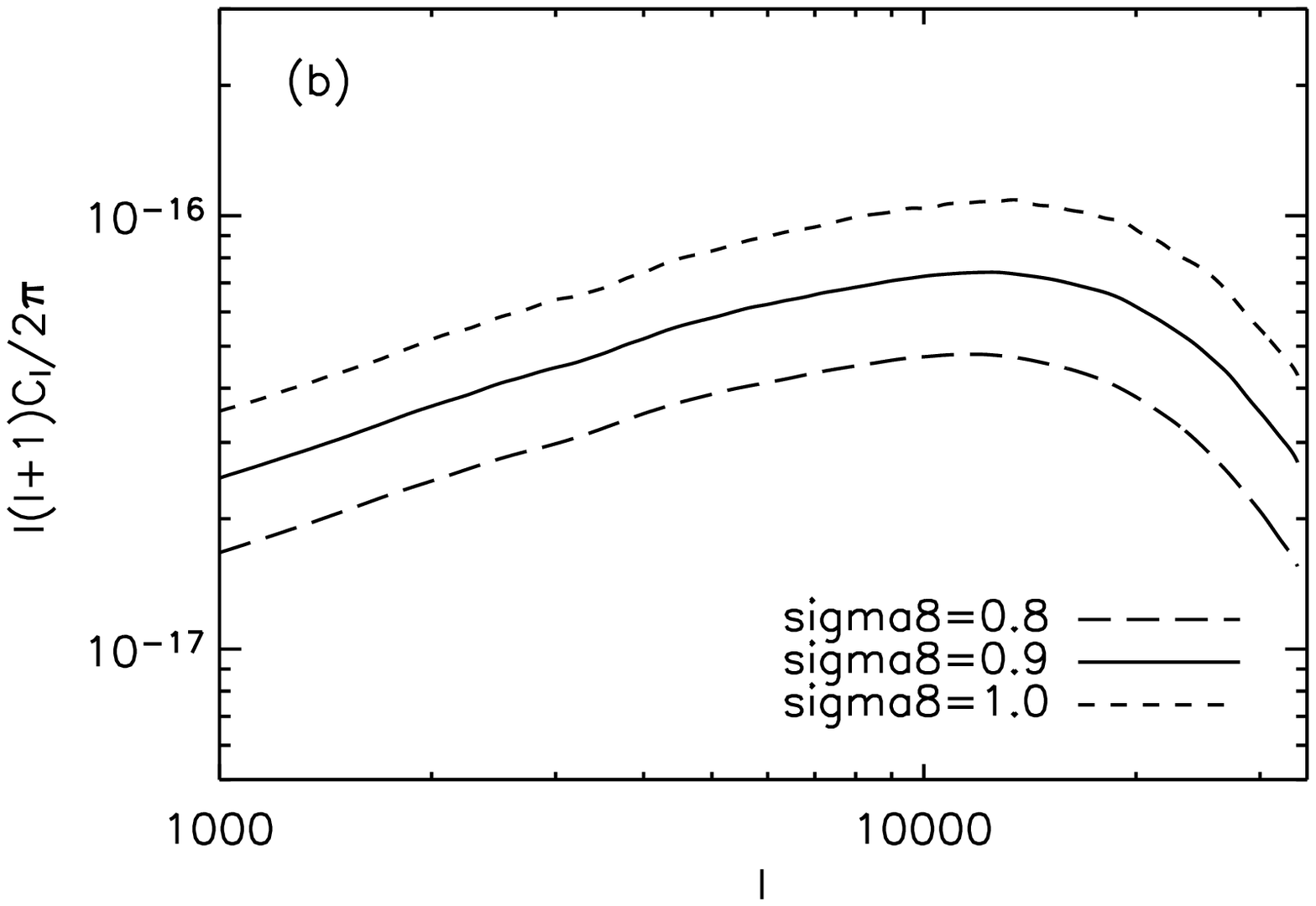}
\includegraphics[angle=0,width=0.5\linewidth]{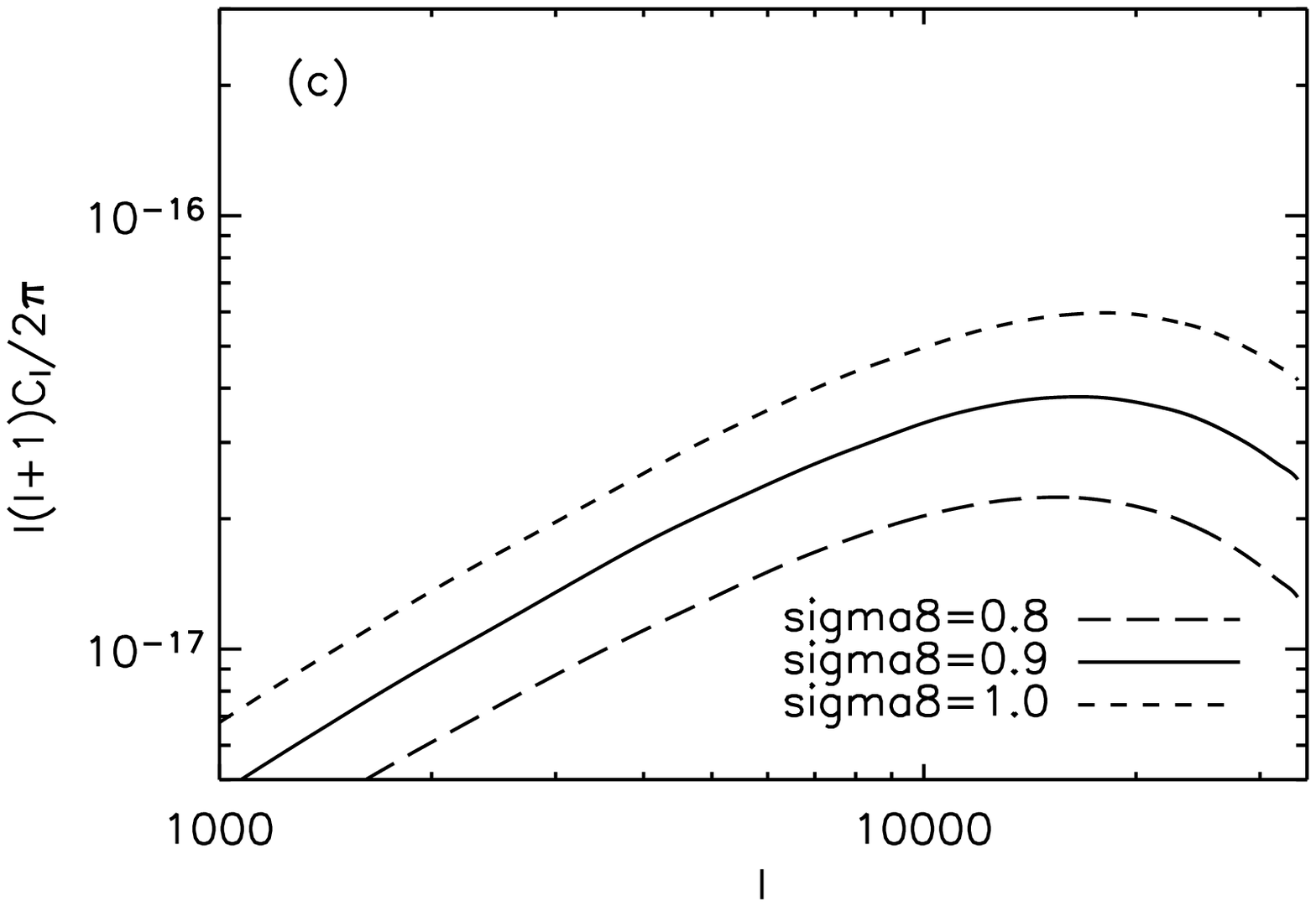}
\caption{Power spectra of secondary polarisation from simulations with
  different $\sigma_8$. Panel (a) is for all ionised particles,
  panel (b) and (c) are for filaments and clusters respectively.}
\label{sigma8}
\end{figure*}

As suggested by Eq.~(\ref{DEL2}), the amplitude of the secondary
polarised signal from clusters is expected to have a strong dependence
on the amplitude of the density fluctuations, which is usually
parameterised by $\sigma_8$.  To estimate the effect of $\sigma_8$ on
our results, we compare in Fig.~(\ref{sigma8}) the power spectra
obtained with the three simulations: $\sigma_8=0.8$ (long-dashed
line), 0.9 (solid line) and 1 (dashed line). Panel (a) shows the total
power spectrum from all ionised particles. Panels (b) and (c) exhibit
the power from the IGM and ICM respectively. Assuming the power
spectra from clusters, filaments and all ionised particles vary like
$C_{E,B l} \propto \sigma_8^n$ we get $n=3.0-4.4$ for all ionised
particles, $3.2-4.0$ for filaments and $3.5-4.6$ for clusters. The
amplitude of the polarisation power spectrum is sensitive with
$\sigma_8$, however, the value of the index $n$ is smaller than that
obtained for the SZ power spectrum variation $6 \lsim n \lsim 9$ (see
for example Zhang, Pen, \& Wang (2002), Bond et al. (2002), Komatsu \&
Seljak (2002))

We compare our secondary polarisation spectra (solid, dotted and
long-dashed lines in Fig.~({\ref{cl_all}})) with the primary $E$ and
$B$ modes sue to scalar and tensor mode perturbations (dashed and
dot-dashed lines respectively).  The $E$ mode signal from clusters and
filaments dominates the primary signal at $\ell \gsim 5000$.  Below
this scale, the secondary polarised signal is always very small and
therefore do not significantly contaminate the primary signal.
Gravitational waves, i.e. tensor mode perturbations, induce primary
$B$ mode polarisation which amplitude depends on the ratio of tensor
to scalar quadrupole, $r=C^{T}_{\ell}/C^{S}_{\ell}$, which is not well
constrained at present.  We compute the primary $B$ mode signal, with
$0<r<0.9$ within the upper limits (at $95\%$ confidence level) set by
WMAP+CBI+ACBAR+2dFGRS+Lyman$\alpha$~\citep{spergel} observations and
the tensor spectral index $n_t=-r/8$, and compare it with our
secondary $B$ mode signal. We conclude that within $0<r<0.9$ the secondary
signal does not prevent us from detecting gravitational wave induced
polarisation at scales $\ell \lsim 100$.

Gravitational lensing by large scale structures modifies slightly the
primary $E$ mode power spectrum. Most noticeably it generates, through
mode coupling, $B$ mode polarisation out of pure $E$ mode signal
(Benabed, Bernardeau, \& van Waerbeke 2001;Zaldarriaga \& Seljak
1998). The gravitational lensing-induced signal, which peaks around
$\ell \sim 1000$, is a major issue for primary $B$ mode
measurement. We compare in Fig.~(\ref{cl_all}) the $B$ mode power
spectrum of the polarised signal from gravitational lensing by large
scale structures to that associated with clusters and filaments. We
find the latter is about two orders of magnitude smaller and it
dominates at very small angular scales ($\ell \gsim 10000$). The
polarised signal from clusters and filaments might be observed at
larger angular scales if the lensing-induced $B$ modes can be
significantly cleaned by appropriate techniques as those proposed for
example by Seljak \& Hirata (2004). We note that the dependence of
lensing-induced polarisation on $\sigma_8$ is different from that of
clusters and filaments. An increase of $\sigma_8$ would enhance more
the latter with respect to the former.

\section{Conclusion}

In this paper, we present the first computation of the $E$ and $B$ mode
quadrupole-induced polarisation power spectra of galaxy clusters and
filaments using hydrodynamical LSS simulations. 

We find that the power spectrum from filaments dominates the power
from clusters by a factor of about two on small scales, and by one
order of magnitude on large scales. The secondary polarisation signal
can thus provide complementary information to those we get from
secondary temperature fluctuations especially for filaments. Assuming
the power spectra vary like $\sigma_8^n$ we find $n$ in the range 3 to
4.6 depending on the gas phase.

As expected, the secondary polarisation power spectra dominate at
small angular scales. Although this first computation of the
polarisation from filaments shows that the signal can be one order of
magnitude larger on large scales, the secondary polarisation is still
subdominant. Therefore it is not a major problem for the primordial
$B$ mode detection.

\acknowledgments 

The authors thank N. Sugiyama for fruitful discussions. GCL
acknowledges support by The National Science Council of Taiwan
NSC92010P. ADS acknowledges support from CMBnet EU TMR network and
JSPS for partial fundings. NA acknowledges NAOJ for support.  The
authors further thank the NAOJ for hospitality during the finalisation
of the paper.




\begin{thebibliography}{0}
\bibitem[Audit \& Simmons 1999]{audit}
 Audit, E., \& Simmons, J. F. L. 1999 MNRAS., {\bf 305}, 27

\bibitem[Bardeen et al. 1986]{bardeen:1986} Bardeen, J.~M., Bond,
J.~R., Kaiser, N., \& Szalay, A.~S., 1986, \apj, {\bf 304}, 15

\bibitem[Baumann, Cooray, \& Kamionkowski 2003]{Baumann}
Baumann, D., Cooray, A., \& Kamionkowski, M., 2003, New Astron., {\bf 8}, 565
\bibitem[Benabed, Bernardeau, \& van Waerbeke 2001]{Benabed}
Benabed, K., Bernardeau, F., \& van Waerbeke, L., 2001, Phys.Rev. {\bf
D63}, 043501
\bibitem[Bennett et al. 1996]{bennett}
Bennett, C. L., et al., 1996, \apj, {\bf 462}, L49
\bibitem[Benoit et al. 2003]{benoit}
Benoit, A., et al., astro-ph/0306222 
\bibitem[Bond et al. 2002]{CBI}
Bond, J. R., et al. 2002, \apj submitted, astro-ph/0205386
\bibitem[Cooray \& Sheth 2002]{Cooray}
Cooray, A., \& Sheth, R., 2002, Phys.Rept. {\bf 372}, 1
\bibitem[Couchman 1991]{couchman:1991}
Couchman, H.~M.~P., 1991, Ap.J, {\bf 368}, L23
\bibitem[Couchnman, Thomas, \& Pearce 1995]{CTP}
Couchnman, H. M. P., Thomas, P. A., Pearce, F. R., 1995, \apj, {\bf 452}, 797
\bibitem[da Silva et al. 2001]{antonio}
 da Silva, A., et al., 2001, \apj, {\bf 561L}, 15D
\bibitem[Eke, Navarro, \& Frenk 1998]{eke:1998}
Eke, V.~R., Navarro, J.~ F., \& Frenk, C.~S., 1998, \apj, {\bf 503}, 569
\bibitem[Hu et al. 1998]{Hu2}
Hu, W., Seljak, U., White, M., \& Zaldarriaga, M., 1998, \PRD, {\bf 57}, 3290

\bibitem[Hu 2000]{Hu}
  Hu, W., 2000, \apj, {\bf 529}, 12

\bibitem[Johnson et al. 2003]{max}
Johnson, B. R. et al., 2003, New Astronomy Reviews, {\bf 47}, 1067J
\bibitem[Kamionkowski \& Loeb 1997]{KL}
Kamionkowski, M., \& Loeb, A., 1997, \PRD, 56, 5411
\bibitem[Kogut et al. 2003]{wmap}
Kogut, A. et al, 2003, Astrophys.J.Suppl. {\bf 148}, 161
\bibitem[Komatsu \& Seljak 2002]{KOMATSU}
Komatsu, E., \& Seljak U., 2002, MNRAS, {\bf 336}, 1256
\bibitem[Lavaux et al. 2004]{Lavaux}
Lavaux, G., Diego, J. M., Mathis, H., \& Silk, J. 2004, MNRAS. {\bf 347}, 729
\bibitem[Leitch et al. 2002]{dasi}
Leitch, E. M. et al., 2002, Nature {\bf 420}, 763

\bibitem[Liu et al. 2001]{Liu}
Liu, G. C., Sugiyama, N. Benson, A. J., Lacey, C. G., \& Nusser, A.,
2001, \apj, {\bf 561}, 504
\bibitem[Lo et al. 2001]{AMIBA}
Lo, K., et al., 2001, AIPC., {\bf 586}, 172L
\bibitem[Monaghan 1992]{monaghan:1992}
Monaghan, J.~J., 1992, Ann. Rev. Astron. Astrophys., {\bf 30}, 543 
\bibitem[Montroy et al. 2003]{boom}
Montroy, T. et al.,  2003, New Astronomy Reviews, {\bf 47}, 1057

\bibitem[ Ng \& Liu 1999]{NL}
 Ng, K.-W., \&  Liu G.-C., 1999, Int. J. Mod. Phys. {\bf D8}, 61-83
\bibitem[Ng \& Ng 1996]{ng}
 Ng, K. L., \& Ng, K.-W., 1996, \apj,  {\bf 456}, 413 
%
\bibitem[Ohno et al. 2002]{Ohno}
 Ohno, H., Takada, M., Dolag, K., Bartelmann, D., \& Sugiyama, N, 2002,
	\apj, {\bf 584}, 599
\bibitem[de Oliveira-Costa et al. 2003] {Oliveira}
de Oliveira-Costa, A., et al. 2003, Phys.Rev. D {\bf 68} 083003

\bibitem[Pearce \& Couchman 1997]{PC}
Pearce, F. R., \& Couchnman, H. M. P., New Astronomy, {\bf 2}, 411(1997)

\bibitem[Portsmouth 2004]{JP}
Portsmouth, J., astro-ph/0402173
\bibitem[Santos et al. 2003]{Santos}
Santos, M. B., Cooray, A., Haiman, Z., Knox, L., \& C.-P. Ma, 2003, \apj,
    {\bf 598}, 756
\bibitem[Sazonov \& Sunyaev 1999]{SS}
Sazonov, S. Y., \& Sunyaev, R. A.,  MNRAS, {\bf 310}, 765(1999)

\bibitem[Seljak \& Hirata 2004] {SH}
Seljak, U., \& Hirata, C. M., 2004, Phys. Rev. D {\bf 69}, 043005

\bibitem[Spergel 2003]{spergel}
Spergel, D.N., et al., 2003, ApJS, {\bf 148}, 175

\bibitem[Sugiyama 1995]{sugiyama:1995}
Sugiyama, N., 1995, Astrophys.J.Suppl. {\bf 100}, 281

\bibitem[Sunyaev \& Zel'dovich 1980]{SZ}
Suntaev, R. A., \& Zel'dovich, Y. B., 1980, MNRAS, {\bf 190}, 413
\bibitem[Takada, Ohno, \& Sugiyama 2001]{takada}
Takada, M., Ohno, H., \& Sugiyama, N. astro-ph/0112412 
\bibitem[Valageas, Schaeffer, \& Silk 2002]{valageas:2002}
Valageas, P. Schaeffer, R., \& Silk, J., 2001, Astron. Astrophys., {\bf 367}, 1
\bibitem[Vishniac 1987]{Vishniac}
 Vishniac, E. T.,1987, ApJ, {\bf 322}, 597

\bibitem[Yurchenko 2002]{planck}
Yurchenko, V., 2002, AIP Conf.Proc. {\bf 616}, 234;
\bibitem[Zaldarriaga \&  Seljak 1997]{ZS}
Zaldarriaga, M., \& Seljak, U., 1997,  Phys. Rev. {\bf D55}, 1830
\bibitem[Zaldarriaga \&  Seljak 1998]{ZS2}
Zaldarriaga, M., \& Seljak, U., 1998, \PRD, {\bf 58}, 023003

\bibitem[Zhang, Pen, \& Wang 2002]{ZHANG}
Zhang, P., Pen, U., \& Wang, B., 2002, \apj, {\bf 577}, 555
\end{thebibliography}
\end{document}